\documentclass[11pt,a4paper]{article}
\usepackage[utf8]{inputenc}
\usepackage[T1]{fontenc}
\usepackage{lmodern}
\usepackage[a4paper,margin=2.5cm]{geometry}
\usepackage{amsmath,amssymb}
\usepackage{mathrsfs}
\usepackage{graphicx}
\usepackage{booktabs}
\usepackage[font=small,labelfont=bf]{caption}
\usepackage[numbers,sort&compress]{natbib}
\usepackage{microtype}
\usepackage{authblk}
\usepackage[colorlinks=true,citecolor=blue,linkcolor=blue,urlcolor=blue]{hyperref}
\graphicspath{{figures/}}
\usepackage{xcolor}

\newcommand{\Lag}{\mathscr{L}}
\newcommand{\Lf}{\mathscr{L}_{F}}
\newcommand{\rp}{r_{+}}
\newcommand{\rmn}{r_{-}}
\newcommand{\dd}{\mathrm{d}}
\newcommand{\Ll}{L_\ell}

\title{\bfseries Regular magnetic black holes with anti-de Sitter cores in scalar--tensor gravity: axial stability and radial instability}
\author{Rustam Ibadov}
\author{Najibullokhon Shukurullokhon}
\affil{Department of Theoretical Physics and Quantum Electronics, Institute of Engineering Physics, Samarkand State University named after Sharof Rashidov, Samarkand 140104, Uzbekistan}

\date{\small E-mail: \href{mailto:ibrustam@samdu.uz}{ibrustam@samdu.uz} (corresponding author)}

\begin{document}
\maketitle

\begin{abstract}
We construct static, asymptotically flat regular black holes in Einstein-frame scalar--tensor gravity, sourced by a canonical scalar coupled to magnetic nonlinear electrodynamics. Prescribing the density and scalar profiles, we reconstruct the metric, the scalar--photon coupling and the potential. Regularity requires the mass function to vanish at the centre, which fixes the ADM mass, and the central $r^2$ growth of the density to be compensated by the gradient energy of the scalar. The solutions have an anti-de Sitter core supported by a negative potential and a nearly Reissner--Nordstr\"om exterior; absence of ghosts restricts the magnetic charge to an interval that shrinks to zero at a finite core depth. Axial perturbations reduce to a symmetric two-channel master system whose positive-definite potential guarantees mode stability; spectral and time-domain quasinormal frequencies agree. Radial perturbations, by contrast, are unstable: the reconstructed potential behaves as $V\propto-\varphi^5$ near the asymptotic vacuum, the scalar becomes tachyonic outside the horizon, and every ghost-free black hole has a growing monopole mode with $\Gamma M\approx0.14$--$0.28$. An angular Laplacian instability develops near the centre; for black holes it may lie entirely within the inner horizon, while horizonless solutions leave it uncovered and are therefore excluded.
\end{abstract}

\section{Introduction}\label{sec:intro}

The theorems of Penrose and Hawking~\citep{Penrose1965,HawkingPenrose1970} tell us that, once the usual energy conditions hold, collapse in General Relativity (GR) ends in a geodesically incomplete spacetime. Regular black holes (RBHs) provide a phenomenological escape: the singularity is replaced by a central region in which the curvature remains finite. The earliest example goes back to Bardeen~\citep{Bardeen1968}; further models were suggested by Dymnikova~\citep{Dymnikova1992} and by Hayward~\citep{Hayward2006}, and the subject is critically reviewed in~\citep{Frolov2016}. A physically motivated source for such geometries is nonlinear electrodynamics (NED)~\citep{AyonBeatoGarcia1998,AyonBeatoGarcia2000,Bronnikov2001,Bronnikov2018}. Bronnikov proved~\citep{Bronnikov2001,Bronnikov2000} that a Lagrangian $L(F)$ with Maxwell behaviour in the weak-field limit admits no static, spherically symmetric solution with a regular centre if the field is purely electric. Magnetic configurations are not affected by this no-go result, provided that $L(F)$ approaches a finite limit for $F\to\infty$. Causality of the NED sector imposes further restrictions: Russo and Townsend proved that every causal NED theory satisfies both the dominant and the strong energy conditions~\citep{RussoTownsend2024}, and that in Einstein--NED theory any causal model gives a metric that is singular at the centre whenever the solution approaches Reissner--Nordstr\"om at infinity~\citep{RussoTownsend2026}. A regular NED core therefore necessarily involves acausal propagation somewhere in its interior. Regular cores also leave imprints on astrophysical observables; in particular, thin accretion disks have been studied around the Bardeen and Hayward geometries~\citep{Akbarieh2025} and around regular black holes supported by NED~\citep{Khoshrangbaf2025}. For the Bardeen and Hayward metrics, the free parameters of the core move the innermost stable circular orbit inwards, which raises the flux, temperature and luminosity of the disk and the radiative efficiency relative to Schwarzschild. The interplay between regularity and energy conditions has also been examined systematically. Static regular metrics of Kerr--Schild form satisfying the weak energy condition form broad families containing the Bardeen, Hayward and Dymnikova metrics as particular members~\citep{WangBattista2026b}, whereas in static, spherically symmetric spacetimes with $g_{tt}g_{rr}\neq-1$ the presence of horizons may be accompanied by violation of the null energy condition~\citep{WangBattista2026a}.

Two further ingredients enlarge the set of possible solutions considerably. One of them is a scalar field. It can interact nonminimally with curvature, which is the defining feature of scalar--tensor theories (STT), or with the electromagnetic invariant via a dilaton-like factor $f(\varphi)$. Scalar fields are able to support objects that NED by itself cannot, for instance wormholes and black universes~\citep{BronnikovFabris2006,BronnikovKonoplyaZhidenko2012,Bronnikov2024,Bronnikov2025}. In combination with non-Abelian gauge fields they generate hairy wormholes that branch off Einstein--Yang--Mills solutions of the Bartnik--McKinnon kind~\citep{Hauser2014}. For Einstein--scalar--Gauss--Bonnet gravity, a scalar self-interaction potential widens the domain of traversable wormholes~\citep{Ibadov2020}, and wormholes carrying NUT charge can be dressed with scalar hair in the same theories~\citep{Ibadov2021,Ibadov2022}. The lesson of these works is that the form of $V(\varphi)$, together with the way $\varphi$ couples to matter, largely controls which nontrivial solutions are allowed. The other ingredient is the inverse, or reconstruction, approach~\citep{BronnikovBadalovIbadov2023,BronnikovRubin2013}. Here the geometry is chosen first, and the potential and couplings needed to support it are found afterwards. Reconstruction of the source has also been used recently to describe how regular black holes may form in generalised Vaidya collapse; there, positivity of the matter sectors was shown to forbid a de Sitter core with nonnegative tangential equations of state~\citep{BattistaOvgunVertogradov2026}, which gives additional motivation for considering cores of a different type. The freedom to prescribe the geometry comes with an obligation: the theory produced in this way must be examined for consistency, in particular for ghosts, and its solutions must be examined for stability~\citep{MorenoSarbach2003,NomuraYoshidaSoda2020,DeFeliceTsujikawa2025PRL,DeFeliceTsujikawa2025,Bronnikov2026a}. For radial (monopole) perturbations, which are usually regarded as the most dangerous ones, a single master equation with an effective potential $V_{\rm eff}$ has been derived for general scalar--electromagnetic couplings $\Lag(\psi,F)$ in the Einstein frame of Bergmann--Wagoner--Nordtvedt STT~\citep{Bronnikov2026a}. That analysis showed that, when the NED reduces to Maxwell theory at weak fields, $V_{\rm eff}$ loses all dependence on the particular NED model as the charge tends to zero, that configurations with $F=0$, i.e.\ with electric and magnetic charges of equal magnitude, inherit the stability properties of their STT--Maxwell counterparts, and that, among scalar--electrovacuum configurations, black holes tend to form islands of stability, whereas solutions with naked singularities are generically unstable.

In an earlier preprint~\citep{IbadovShukurullokhon2026} we studied regular solutions with AdS-type cores in this setting. There, however, the scalar potential was fixed from the outset, the field profiles were chosen as analytic ans\"atze, and the axial perturbations were handled only approximately. The approach taken below is different. We derive the potential and the coupling instead of postulating them, so that the conditions of regularity and of ghost freedom become explicit, and we formulate the axial problem as an exact two-channel master system.

Concretely, we obtain magnetically charged regular black holes whose central region is of \emph{anti-de Sitter} (AdS) type, in contrast with the de Sitter core that is usual for NED-supported RBHs; the change is due to the scalar field. Regular objects with AdS cores have recently appeared in a quite different setting as well, namely as polymerised vacuum solutions inspired by loop quantum gravity, where both black holes and horizonless ultracompact configurations are fixed uniquely by the mass~\citep{LiuSoranidis2026}; in our case the AdS core is instead supported by a classical matter source. No phantom is needed: $\varphi$ has a standard kinetic term, and the central energy density turns negative because the potential does. Our main findings can be summarised as follows.

\begin{enumerate}
\item[(i)] We point out two regularity requirements that no inverse construction can avoid. First, $m(0)=0$, and as a consequence the ADM mass cannot be chosen independently, being fixed by the charges and the core radius. Second, $f(\varphi_c)$ stays finite and nonzero only if the $r^2$ term of the central energy density equals the gradient energy of the scalar, $\rho_2=8\varphi_2^2$. Both requirements are met in closed form by an elementary family depending on two parameters.
\item[(ii)] The no-ghost condition for the photon, $f>0$, is a real restriction. Black holes without ghosts exist only for magnetic charges inside a certain interval, and this interval disappears once the AdS core becomes deep enough.
\item[(iii)] Allowing a general scalar-dependent NED Lagrangian $\Lag(F,\varphi)$ on a magnetically charged background, we reduce the coupled problem of odd-parity (axial) metric and even-parity electromagnetic perturbations to master form. We test these equations numerically against the full set of linearised field equations and use the positivity of the potential matrix, checked over the whole ghost-free window, to establish axial mode stability.
\item[(iv)] We compute the lowest axial quasinormal modes (QNMs) by a spectral method and confirm them with time-domain simulations. For $\ell=2$ the branch dominated by the electromagnetic field departs from its Reissner--Nordstr\"om (RN) analogue by 10--18\% in ${\rm Re}\,\omega$, so it is the most sensitive diagnostic of the core.
\item[(v)] Following De Felice and Tsujikawa~\citep{DeFeliceTsujikawa2025PRL,DeFeliceTsujikawa2025}, we find the central zone in which the complementary (polar-gravity) sector suffers a Laplacian instability along the angular directions. In black holes this region can be hidden behind the inner horizon; in horizonless objects this is impossible.
\item[(vi)] Using the general effective potential for monopole perturbations of Ref.~\citep{Bronnikov2026a}, we show that all ghost-free black holes of the family are unstable against radial perturbations, with $e$-folding times of $4$--$7\,M$. The instability is caused by the negative curvature of the reconstructed potential near $\varphi=0$. It points to a third requirement on inverse constructions, in addition to regularity and ghost freedom: the effective scalar mass outside the horizon should be nonnegative.
\end{enumerate}

The paper is organised as follows. Section~\ref{sec:setup} introduces the action, the transformation between frames and the field equations. In Section~\ref{sec:static} we specialise to static spherical symmetry, formulate the inverse problem and derive the conditions of regularity. The model itself is constructed and analysed in Section~\ref{sec:model}. Section~\ref{sec:axial} is devoted to axial perturbations and QNMs, Section~\ref{sec:radial} to radial perturbations, and Section~\ref{sec:polar} to the remaining perturbation sector and to thermodynamics. Our conclusions are collected in Section~\ref{sec:concl}, while the appendices contain the derivation of the master system and details of the numerical work. We use the signature $(-,+,+,+)$ and units with $\kappa^2=8\pi G$ and $c=1$.

\section{Action and equations of motion}\label{sec:setup}

\subsection{From the Jordan to the Einstein frame}

We begin with a general STT in which the NED sector is also coupled nonminimally,
\begin{equation}
S_J=\int \dd^4x\sqrt{-\tilde g}\left[\mathscr{F}(\Phi)\tilde R-\frac12 h(\Phi)\tilde g^{\mu\nu}\partial_\mu\Phi\partial_\nu\Phi-\tilde V(\Phi)+\tilde{\Lag}(\tilde F,\Phi)\right],
\label{eq:SJ}
\end{equation}
with $\tilde F=\tilde F_{\mu\nu}\tilde F^{\mu\nu}$. The conformal transformation $g_{\mu\nu}=\Omega^2\tilde g_{\mu\nu}$, $\Omega^2=2\kappa^2\mathscr{F}(\Phi)$, does not act on the gauge potential or on $F_{\mu\nu}$, while $\tilde F=\Omega^4F$ and $\sqrt{-\tilde g}=\Omega^{-4}\sqrt{-g}$. Inserting $\tilde R=\Omega^2[R+6\Box\ln\Omega-6(\partial\ln\Omega)^2]$ and discarding a total derivative, we find the action in the Einstein frame,
\begin{equation}
S=\int\dd^4x\sqrt{-g}\left[\frac{R}{2\kappa^2}-\frac12(\partial\varphi)^2-V(\varphi)+\Lag(F,\varphi)\right],
\label{eq:SE}
\end{equation}
in which
\begin{equation}
\left(\frac{\dd\varphi}{\dd\Phi}\right)^2=\frac{h}{\Omega^2}+\frac{6}{\kappa^2}\left(\frac{\dd\ln\Omega}{\dd\Phi}\right)^2,\qquad
V=\frac{\tilde V}{\Omega^4},\qquad \Lag(F,\varphi)=\frac{\tilde\Lag(\Omega^4F,\Phi)}{\Omega^4}.
\label{eq:map}
\end{equation}
The Einstein-frame scalar has a normal (non-phantom) kinetic term whenever the first relation in Eq.~\eqref{eq:map} has a positive right-hand side. As long as $\Omega$ stays smooth and positive, the map between the two frames is nonsingular; regularity and linear stability proved in the Einstein frame therefore carry over to the Jordan frame~\citep{Bronnikov1973,Bronnikov2024}. Strictly, this requires $\Omega$ to remain smooth and positive also at the horizon and at spatial infinity, because the boundary conditions for perturbations should be imposed in the physically preferred Jordan frame; when $\Omega$ degenerates somewhere, stability conclusions can differ between particular STTs sharing the same Einstein frame~\citep{Bronnikov2026a}. From here on we work with the action~\eqref{eq:SE} and assume a factorised coupling,
\begin{equation}
\Lag(F,\varphi)=f(\varphi)\,L(F),\qquad L(F)\xrightarrow[F\to0]{}-\frac F4 ,
\label{eq:fact}
\end{equation}
which ensures Maxwell behaviour wherever $f\to1$.

\subsection{Equations of motion}

Varying the action~\eqref{eq:SE}, we get
\begin{align}
G_{\mu\nu}&=\kappa^2T_{\mu\nu},\label{eq:Einstein}\\
T_{\mu\nu}&=\partial_\mu\varphi\partial_\nu\varphi-g_{\mu\nu}\left[\tfrac12(\partial\varphi)^2+V\right]-4\Lf F_{\mu\alpha}F_\nu{}^\alpha+g_{\mu\nu}\Lag,\nonumber\\
0&=\Box\varphi-V_{,\varphi}+\Lag_{,\varphi},\qquad 0=\nabla_\mu\!\left(\Lf F^{\mu\nu}\right),\label{eq:scalarMaxwell}
\end{align}
where $\Lf\equiv\partial\Lag/\partial F=f(\varphi)L_F$. The sign of the photon kinetic term is correct only if $\Lf<0$. For the models used below $L_F<0$, so this requirement reduces to $f(\varphi)>0$.

As the NED model we take the rational Lagrangian of Kruglov~\citep{Kruglov2014,Kruglov2021,Kruglov2023},
\begin{equation}
L(F)=-\frac{F}{4+\sigma F},\qquad \sigma>0,
\label{eq:Kruglov}
\end{equation}
which behaves as $-F/4$ for $F\to0$ and tends to $L_c=-1/\sigma$ for $F\to\infty$. The Born--Infeld Lagrangian~\citep{BornInfeld1934}, on the other hand, increases like $\sqrt F$ for large $F$, so a magnetic solution built on it cannot have a regular centre~\citep{Bronnikov2001}.

\section{Static, spherically symmetric setting}\label{sec:static}

\subsection{Reduced equations and the reconstruction scheme}

We write the line element as
\begin{equation}
\dd s^2=-e^{2\nu(r)}\dd t^2+e^{2\lambda(r)}\dd r^2+r^2\dd\Omega^2,\qquad e^{-2\lambda}\equiv1-\frac{2m(r)}{r},
\label{eq:metric}
\end{equation}
and take $\varphi=\varphi(r)$ together with the monopole potential $A_\mu\dd x^\mu=-q\cos\theta\,\dd\phi$. Then $F_{\theta\phi}=q\sin\theta$, $F=2q^2/r^4$, and the Maxwell equations in~\eqref{eq:scalarMaxwell} are fulfilled identically. The Einstein tensor has the nonvanishing components
\begin{align}
G^t{}_t&=e^{-2\lambda}\left(\frac1{r^2}-\frac{2\lambda'}{r}\right)-\frac1{r^2}=-\frac{2m'}{r^2},\qquad
G^r{}_r=e^{-2\lambda}\left(\frac1{r^2}+\frac{2\nu'}{r}\right)-\frac1{r^2},\label{eq:Gtt}\\
G^\theta{}_\theta&=G^\phi{}_\phi=e^{-2\lambda}\left(\nu''+\nu'^2-\nu'\lambda'+\frac{\nu'-\lambda'}{r}\right).\label{eq:Gthth}
\end{align}
It is convenient to regard the matter as an anisotropic fluid, $T^\mu{}_\nu={\rm diag}(-\rho,p_r,p_\perp,p_\perp)$, with
\begin{equation}
\begin{gathered}
\rho=\tfrac12e^{-2\lambda}\varphi'^2+V-\Lag,\qquad
p_r=\tfrac12e^{-2\lambda}\varphi'^2-V+\Lag,\\
p_\perp=-\tfrac12e^{-2\lambda}\varphi'^2-V+\Lag-2F\Lf .
\end{gathered}
\label{eq:fluid}
\end{equation}
In this language the equations $G^\mu{}_\nu=\kappa^2T^\mu{}_\nu$ take the form
\begin{align}
m'&=\frac{\kappa^2}{2}r^2\rho,\label{eq:mprime}\\
\nu'+\lambda'&=\frac{\kappa^2}{2}\,r\varphi'^2\ \ge0,\label{eq:nulam}\\
p_\perp&=p_r+\frac r2\left[p_r'+(\rho+p_r)\nu'\right],\qquad \nu'=\frac{m+\kappa^2r^3p_r/2}{r(r-2m)}.\label{eq:TOV}
\end{align}
Eq.~\eqref{eq:TOV} expresses the conservation law $\nabla_\mu T^\mu{}_r=0$. Eq.~\eqref{eq:fluid} also implies the identities
\begin{equation}
\rho+p_r=e^{-2\lambda}\varphi'^2,\qquad \rho+p_\perp=-2F\Lf .
\label{eq:identities}
\end{equation}
Since $\nu'+\lambda'\ge0$ and $\nu+\lambda\to0$ at infinity, we have $\nu+\lambda\le0$ everywhere. Any nontrivial scalar therefore spoils the property $g_{tt}g_{rr}=-1$, which holds for RBHs supported by NED in GR.

Eqs.~\eqref{eq:mprime}--\eqref{eq:identities} define the inverse problem~\citep{BronnikovBadalovIbadov2023}. For given $\rho(r)$, $\varphi(r)$ and $L(F)$, Eq.~\eqref{eq:mprime} yields $m(r)$ and Eq.~\eqref{eq:nulam} yields $\nu+\lambda$; the radial pressure equals $p_r=-\rho+e^{-2\lambda}\varphi'^2$, and $p_\perp$ is obtained from Eq.~\eqref{eq:TOV}. The coupling and the potential then follow without solving any further differential equation,
\begin{equation}
f=\frac{\Lf}{L_F(F)}=-\frac{\rho+p_\perp}{2FL_F(F)},\qquad V=\rho-\tfrac12e^{-2\lambda}\varphi'^2+f\,L(F),
\label{eq:reconstruct}
\end{equation}
and, since $\varphi(r)$ is monotonic, they can be viewed as functions $f(\varphi)$ and $V(\varphi)$. The contracted Bianchi identity then ensures that the scalar equation in~\eqref{eq:scalarMaxwell} is satisfied automatically.

\subsection{Conditions for a regular centre}\label{sec:regularity}

A regular centre needs $m(r)=\tfrac{\kappa^2}{6}\rho_c r^3+O(r^5)$. Integrating Eq.~\eqref{eq:mprime} outwards from $r=0$, we then obtain
\begin{equation}
M\equiv\lim_{r\to\infty}m(r)=\frac{\kappa^2}{2}\int_0^\infty\rho\,r^2\dd r .
\label{eq:Mfix}
\end{equation}
Consequently, after $\rho(r)$ has been chosen, the ADM mass is no longer independent: it is determined by the parameters that enter $\rho$. With the expansions $\varphi=\varphi_c+\varphi_2r^2+O(r^4)$ and $\rho=\rho_c+\rho_2r^2+O(r^4)$, the metric near the centre becomes
\begin{equation}
e^{-2\lambda}=1-\frac{\Lambda_{\rm eff}}{3}r^2+O(r^4),\qquad \Lambda_{\rm eff}=\kappa^2\rho_c=\kappa^2\left[V(\varphi_c)-f(\varphi_c)L_c\right],
\label{eq:Lambda}
\end{equation}
and the curvature invariants remain bounded, with $R\to4\Lambda_{\rm eff}$ and $R_{\mu\nu\alpha\beta}R^{\mu\nu\alpha\beta}\to\frac83\Lambda_{\rm eff}^2$. A de Sitter core corresponds to $\rho_c>0$, an AdS core to $\rho_c<0$.

The central behaviour of the coupling imposes one more condition. For the Kruglov Lagrangian $FL_F=-4F/(4+\sigma F)^2=O(r^4)$ as $r\to0$, whereas the expansion of Eq.~\eqref{eq:TOV} gives
\begin{equation}
\rho+p_\perp=\left(8\varphi_2^2-\rho_2\right)r^2+O(r^4).
\label{eq:rhopp}
\end{equation}
Eq.~\eqref{eq:reconstruct} thus leads to a finite and nonzero $f(\varphi_c)$ only when
\begin{equation}
\rho_2=8\varphi_2^2 .
\label{eq:balance}
\end{equation}
Put differently, the quadratic rise of the density close to the centre must come entirely from the gradient energy of the scalar; if this is not the case, $f\propto r^{-2}$ for $r\to0$.

\subsection{Behaviour at large radii}

Suppose that the scalar is massless with charge $Q$, $\varphi=Q/r+O(r^{-2})$, and that at large distances the NED is in the Maxwell regime, $\Lag\simeq-F/4$. Then $\rho\simeq(Q^2+q^2)/(2r^4)$, and Eqs.~\eqref{eq:mprime} and~\eqref{eq:nulam} give
\begin{equation}
e^{-2\lambda}=1-\frac{2M}{r}+\frac{\kappa^2(q^2+Q^2)}{2r^2}+O(r^{-3}),\qquad
e^{2\nu}=1-\frac{2M}{r}+\frac{\kappa^2q^2}{2r^2}+O(r^{-3}).
\label{eq:asympt}
\end{equation}
At this order the scalar charge is present in $g_{rr}$ whereas $g_{tt}$ retains the RN form, because the scalar contribution drops out there. In the units used below, the magnetic RN metric is $1-2M/r+\kappa^2q^2/(2r^2)$, and it is extremal for $q=\sqrt2\,M/\kappa$.

\section{Explicit ghost-free solutions with AdS interiors}\label{sec:model}

\subsection{Choice of profiles and the resulting metric}

Our choice is
\begin{equation}
\rho(r)=\frac{C}{2}\,\frac{r^2-\beta a^2}{(r^2+a^2)^3},\qquad \varphi(r)=\frac{Q}{\sqrt{r^2+a^2}},
\label{eq:ansatz}
\end{equation}
where the core radius is set by $a>0$ and its depth by $0\le\beta<1$. As $\rho\to C/(2r^4)$ for large $r$, comparison with Eq.~\eqref{eq:asympt} gives $C=q^2+Q^2$. At small $r$ one finds $\rho_c=-C\beta/(2a^4)\le0$, $\rho_2=C(1+3\beta)/(2a^6)$ and $\varphi_2=-Q/(2a^3)$. Imposing the balance condition~\eqref{eq:balance} determines how $C$ splits into its scalar and magnetic parts:
\begin{equation}
Q^2=\frac{C(1+3\beta)}{4},\qquad q^2=\frac{3C(1-\beta)}{4}.
\label{eq:charges}
\end{equation}
Closed-form expressions follow for $m(r)$ and for the metric:
\begin{align}
m(r)&=\frac{\kappa^2C}{32a}\left\{(3-\beta)\arctan\frac ra-\frac{ar\left[(3-\beta)a^2+(5+\beta)r^2\right]}{(r^2+a^2)^2}\right\},\label{eq:mass}\\
M&=\frac{\pi\kappa^2C(3-\beta)}{64\,a},\label{eq:Mvalue}\\
\nu+\lambda&=-\frac{\kappa^2Q^2(2r^2+a^2)}{8(r^2+a^2)^2}.\label{eq:nuplam}
\end{align}
After setting $\kappa^2=1$ and $M=1$, the family is labelled by the two parameters $(q,\beta)$, with
\begin{equation}
C=\frac{4q^2}{3(1-\beta)},\qquad a=\frac{\pi C(3-\beta)}{64},
\end{equation}
and $Q$ given by Eq.~\eqref{eq:charges}. The core is flat (Minkowski) when $\beta=0$ and AdS when $\beta>0$. Pushing the expansion of Eq.~\eqref{eq:TOV} to the next order, we find $\rho+p_\perp=K_4r^4+O(r^6)$ with
\begin{equation}
K_4=\frac{C\left[36a^2(1-\beta)+7C\beta(1+3\beta)\right]}{48\,a^{10}}>0,
\label{eq:K4}
\end{equation}
and Eq.~\eqref{eq:reconstruct} gives $f(\varphi_c)=K_4\sigma^2q^2/4$.

Three properties of this construction are worth stressing:
\begin{enumerate}
\item[(a)] The geometry is fixed by $(q,\beta)$ alone and does not feel the NED scale $\sigma$. The same is true of the horizons, of the curvature invariants and, as Section~\ref{sec:axial} will show, of the axial perturbation potential. The scale $\sigma$ enters only the reconstructed $f(\varphi)$ and $V(\varphi)$. We fix it through the requirement $f(\varphi_c)=1$, i.e.\ $\sigma=\sigma_1\equiv2/(q\sqrt{K_4})$; then $f$ equals unity at $r=0$ as well as for $r\to\infty$. Other values of $\sigma$ are discussed in Section~\ref{sec:polar}.
\item[(b)] At large distances the reconstruction gives $f=1+O(\varphi)$ and $V=O(\varphi^5)$. Hence the scalar is massless, as required by its $1/r$ decay and by Eq.~\eqref{eq:asympt}. Two features of this behaviour deserve comment. Since $f'(0)\neq0$, the magnetic field always sources the scalar, so the RN metric is not a solution of the reconstructed theory and every magnetically charged configuration carries scalar hair. Moreover, $V\simeq-c\,\varphi^5$ with $c>0$ near $\varphi=0$, and $V$ reaches large negative values in the core; the asymptotic vacuum $\varphi=0$ is therefore not the global minimum of $V$. Its possible nonperturbative decay lies outside the classical analysis of this paper; the negative curvature of $V$ near $\varphi=0$ has, however, a direct perturbative consequence, analysed in Section~\ref{sec:radial}.
\item[(c)] The negative central density is produced by the scalar potential, $V(\varphi_c)=\rho_c-f(\varphi_c)/\sigma<0$. The scalar kinetic term is canonical at all radii, so no phantom field is involved.
\end{enumerate}
We stress that $f(\varphi)$ and $V(\varphi)$ depend on $(q,\beta,\sigma)$. Each point of the parameter space thus defines its own theory, in which the black-hole mass is fixed by Eq.~\eqref{eq:Mvalue}; the two-parameter family is a family of theories, each admitting the solution~\eqref{eq:ansatz}, rather than a continuous family of solutions within one theory. This is a generic property of inverse constructions, and it has to be kept in mind when quantities are compared across the family, for instance in Fig.~\ref{fig:temp}.

To check the construction, we substituted the reconstructed $\nu,\lambda,f,V$ into the complete system~\eqref{eq:Einstein}--\eqref{eq:scalarMaxwell}. All components, the scalar equation among them, are fulfilled with residuals below $3\times10^{-12}$ (Appendix~\ref{app:numerics}).

\subsection{Horizon structure and admissible charges}

\begin{figure}[htbp]
\centering
\includegraphics[width=0.72\textwidth]{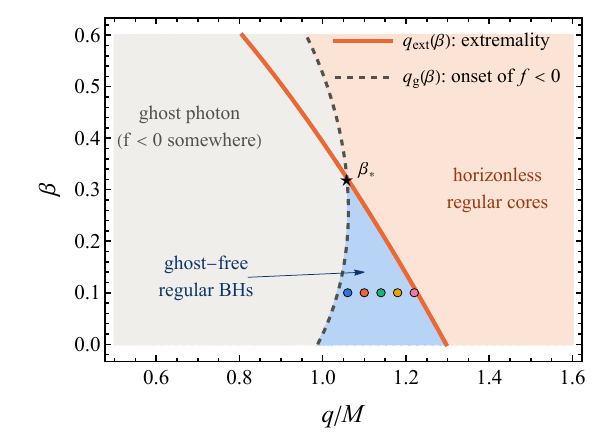}
\caption{Parameter plane $(q,\beta)$ for $M=1$, $\kappa^2=1$. The solid line marks extremality, $q_{\rm ext}(\beta)$, and the dashed line the appearance of $f<0$, $q_{\rm g}(\beta)$. Blue: regular black holes with two horizons and no ghosts. Orange: regular solutions without horizons (Sect.~\ref{sec:polar} discusses their instability). Grey: the coupling becomes negative somewhere. The star shows $\beta_*$, where the ghost-free window closes, and the dots show the fiducial configurations at $\beta=0.1$.}
\label{fig:params}
\end{figure}

Horizons are located at the zeros of $e^{-2\lambda}$. For fixed $\beta$ there are two horizons $\rmn<\rp$ if $q<q_{\rm ext}(\beta)$; at $q=q_{\rm ext}$ they merge, and for larger $q$ no horizon forms. The decisive physical restriction is the absence of ghosts, $f>0$, which according to Eq.~\eqref{eq:identities} is equivalent to $\rho+p_\perp>0$ for all $r$. Below $q_{\rm g}(\beta)$ the inequality fails in a layer at $r\sim a$ near the maximum of $\rho$. Equation~\eqref{eq:balance} matches $\rho$ and $\varphi'^2$ only at leading order, and in this layer the higher-order growth of $\rho$ is no longer compensated by the kinetic energy of $\varphi$. The phase diagram obtained in this way is shown in Fig.~\ref{fig:params}. Regular black holes without ghosts occupy the window
\begin{equation}
q_{\rm g}(\beta)<q<q_{\rm ext}(\beta),
\end{equation}
which contracts to a single point at $\beta_*\simeq0.32$, where $q_{\rm g}=q_{\rm ext}\simeq1.06$. Representative pairs are $(q_{\rm g},q_{\rm ext})=(0.988,1.297)$ at $\beta=0$, $(1.036,1.227)$ at $\beta=0.1$, $(1.059,1.154)$ at $\beta=0.2$ and $(1.060,1.076)$ at $\beta=0.3$. For comparison, the extremal RN solution has $q=\sqrt2\simeq1.414$ in the same units. When $\beta>\beta_*$, only horizonless configurations can be free of ghosts.

\begin{figure}[htbp]
\centering
\includegraphics[width=0.72\textwidth]{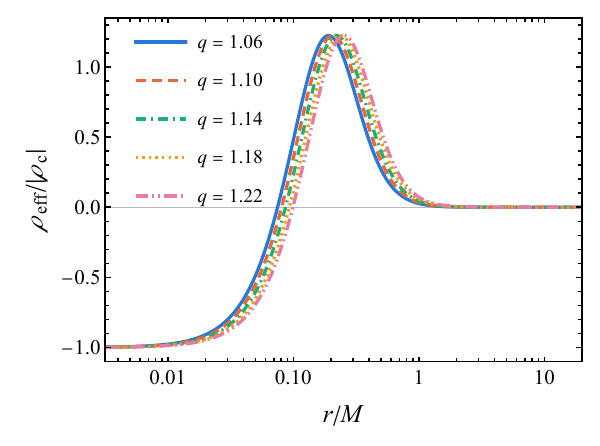}
\caption{Energy density $\rho/|\rho_c|$ for $\beta=0.1$ and the charges of Table~\ref{tab:background}. Each value of $q$ is drawn with the same colour and line style in all figures.}
\label{fig:density}
\end{figure}

\begin{figure}[htbp]
\centering
\includegraphics[width=0.72\textwidth]{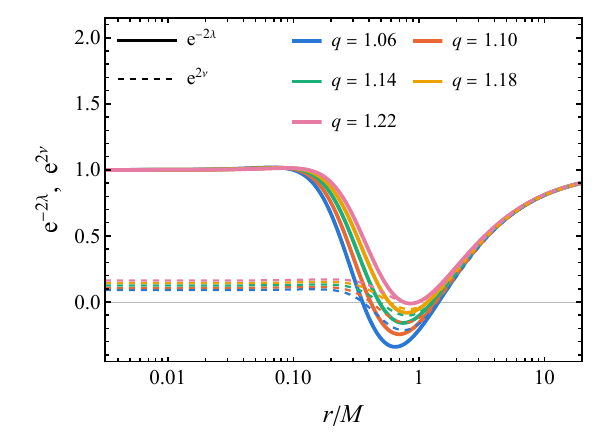}
\caption{The metric functions $e^{-2\lambda}$ (solid) and $e^{2\nu}$ (dashed) for $\beta=0.1$ and the charges of Table~\ref{tab:background}.}
\label{fig:metric}
\end{figure}

\begin{figure}[htbp]
\centering
\includegraphics[width=0.72\textwidth]{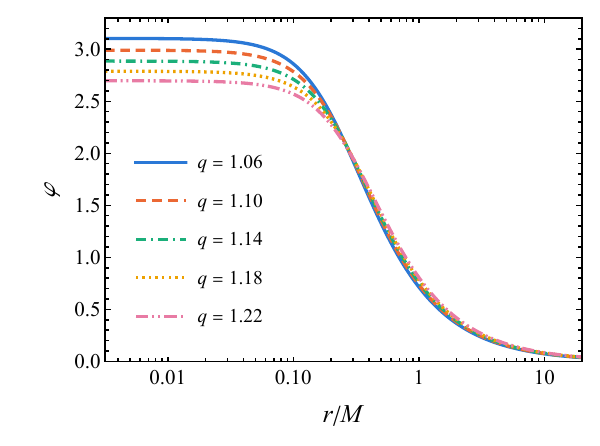}
\caption{Scalar field $\varphi(r)$ for $\beta=0.1$ and the charges of Table~\ref{tab:background}.}
\label{fig:scalar}
\end{figure}

\begin{figure}[htbp]
\centering
\includegraphics[width=0.72\textwidth]{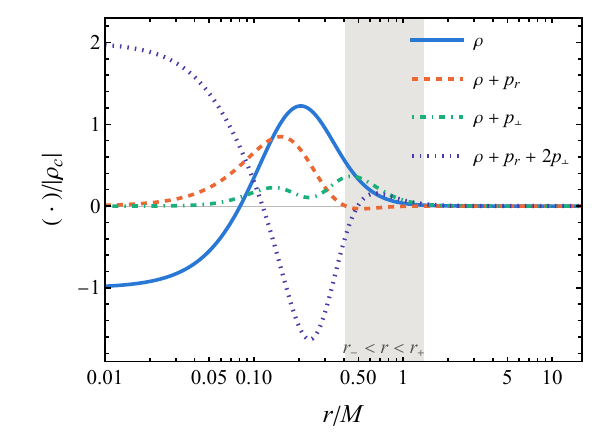}
\caption{Combinations that enter the energy conditions, normalised to $|\rho_c|$, for $\beta=0.1$ and $q=1.10$. Shading marks the interval $\rmn<r<\rp$. Since $\rho+p_\perp=-2F\Lf$ stays positive everywhere, the no-ghost condition $f>0$ is fulfilled.}
\label{fig:EC}
\end{figure}

\begin{figure*}[htbp]
\centering
\includegraphics[width=\textwidth]{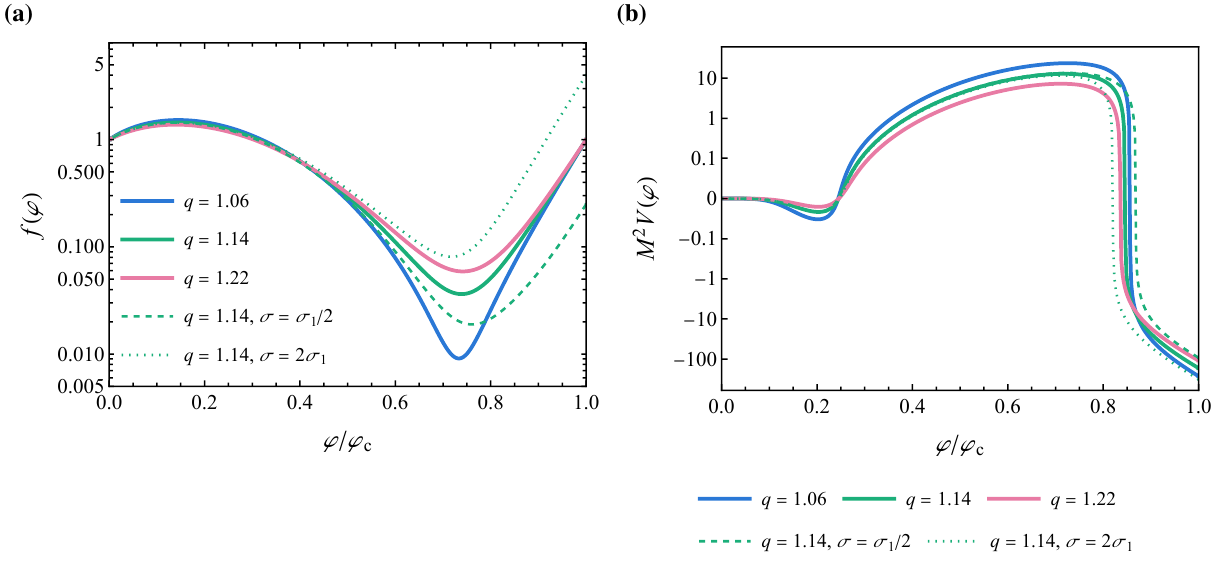}
\caption{Reconstructed functions: (a) the coupling $f(\varphi)$ between scalar and photon and (b) the self-interaction potential $V(\varphi)$, obtained by reconstruction for $\beta=0.1$ and $\sigma=\sigma_1$ (solid lines). The dashed and dotted lines show $\sigma=\sigma_1/2$ and $2\sigma_1$ for $q=1.14$; the metric does not change with $\sigma$. In panel (b) the vertical axis is linear near zero and logarithmic for large $|V|$.}
\label{fig:fV}
\end{figure*}

\begin{figure}[htbp]
\centering
\includegraphics[width=0.72\textwidth]{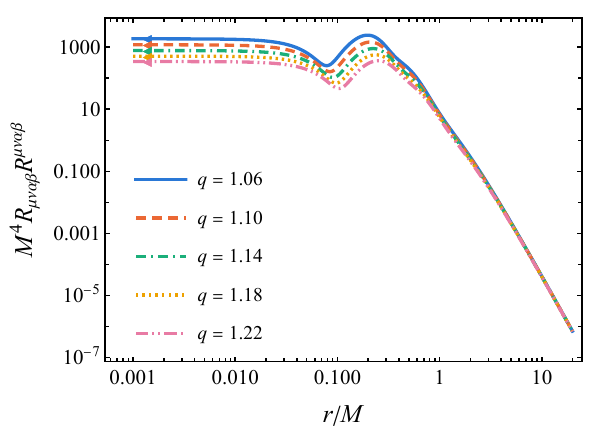}
\caption{Radial behaviour of $R_{\mu\nu\alpha\beta}R^{\mu\nu\alpha\beta}$ at $\beta=0.1$. Triangles on the vertical axis show its exact limit $\frac83\Lambda_{\rm eff}^2$ at $r=0$.}
\label{fig:kretsch}
\end{figure}

Table~\ref{tab:background} collects the fiducial configurations at $\beta=0.1$, and their profiles are shown in Figs.~\ref{fig:density}--\ref{fig:scalar}. Figure~\ref{fig:metric} shows that the region between the horizons, where $e^{-2\lambda}<0$, becomes narrower for larger $q$ and vanishes when $\rmn$ and $\rp$ coalesce at $q_{\rm ext}$. For $r<\sqrt\beta\,a$ the density is below zero; it reaches its maximum near $r\sim a$ and decays as $r^{-4}$ farther out. Because $\nu+\lambda<0$, the lapse at the centre is strongly redshifted, $e^{2\nu(0)}=\exp[-\kappa^2Q^2/(4a^2)]$. The energy conditions are illustrated in Fig.~\ref{fig:EC}. The weak energy condition is violated for $r<\sqrt\beta\,a$, which lies well inside the inner horizon. As expected for an AdS core, the strong energy condition is \emph{satisfied} at the centre, where $\rho+p_r+2p_\perp\to-2\rho_c>0$; its violation is confined to a layer surrounding the peak of $\rho$ ($0.1\lesssim r/M\lesssim0.5$ at $\beta=0.1$), located entirely under the event horizon. In the exterior region the null, weak and strong conditions are all respected. The negative values of $\rho+p_r$ between the horizons seen in Fig.~\ref{fig:EC} do not mean that the null energy condition is violated: in this T-region the coordinate $r$ is timelike, and for any null vector $T_{\mu\nu}k^\mu k^\nu=(k^r)^2\varphi'^2\ge0$, as for every canonical scalar. This point deserves attention because our metric has $g_{tt}g_{rr}\neq-1$, and in such geometries horizons may be accompanied by violation of the null energy condition~\citep{WangBattista2026a}. Here this cannot happen: Eq.~\eqref{eq:nulam} gives $(\nu+\lambda)'=\kappa^2r\varphi'^2/2\ge0$, which in the R-regions is exactly the radial condition $\rho+p_r\ge0$. The dominant energy condition is violated within a narrow shell adjacent to $\rp$, where $p_\perp$ exceeds $\rho$ by up to 25\% close to the lower edge of the window ($q=1.04$) and by 2\% at $q=1.22$. The reconstructed $f(\varphi)$ and $V(\varphi)$ are shown in Fig.~\ref{fig:fV}. When $q$ approaches $q_{\rm g}$, the minimum of $f$ goes to zero, which marks the edge of the ghost-free window. Near this edge the effective electromagnetic coupling, which scales as $1/f$, becomes large (at $q=1.06$ the minimum of $f$ is about $10^{-2}$), so configurations close to $q_{\rm g}$ should be regarded with caution. All curvature invariants are finite: at $r=0$ the Ricci scalar tends to $4\Lambda_{\rm eff}$ and the Kretschmann scalar to $\frac83\Lambda_{\rm eff}^2$ (Fig.~\ref{fig:kretsch}).

\begin{table}[t]
\centering
\caption{Ghost-free fiducial black holes at $\beta=0.1$ ($M=1$, $\kappa^2=1$). The columns give the core scale $a$, the scalar charge $Q$, the central density $\rho_c$, the radii of the inner and outer horizons, the Hawking temperature, the value $\sigma_1$ for which $f(\varphi_c)=1$, the radius $r_L$ of the Laplacian-unstable region at $\sigma=\sigma_1$, and the maximal $\sigma$ for which $r_L<r_-$.}
\label{tab:background}
\small\setlength{\tabcolsep}{4pt}
\begin{tabular}{cccccccccc}
\toprule
$q$ & $a$ & $Q$ & $\rho_c$ & $\rmn$ & $\rp$ & $10^2\,T_H$ & $10^3\sigma_1$ & $r_L$ & $10^3\sigma_{\max}$\\
\midrule
1.06 & 0.2370 & 0.7355 & -26.40 & 0.3558 & 1.4363 & 3.271 & 4.15 & 0.289 & 9.5\\
1.10 & 0.2552 & 0.7633 & -21.14 & 0.4126 & 1.3670 & 3.074 & 5.26 & 0.313 & 16.0\\
1.14 & 0.2741 & 0.7910 & -17.06 & 0.4834 & 1.2834 & 2.786 & 6.62 & 0.337 & 28.0\\
1.18 & 0.2936 & 0.8188 & -13.87 & 0.5793 & 1.1746 & 2.308 & 8.26 & 0.362 & 53.9\\
1.22 & 0.3139 & 0.8465 & -11.36 & 0.7547 & 0.9862 & 1.100 & 10.22 & 0.389 & 145.3\\
\bottomrule
\end{tabular}
\end{table}

\section{Odd-parity gravitational perturbations}\label{sec:axial}

\subsection{Reduction to two coupled channels}

We decompose the perturbations into spherical harmonics~\citep{ReggeWheeler1957,Zerilli1970}. When the background field is magnetic rather than electric, the electromagnetic perturbation that mixes with a given metric parity is the one of opposite parity~\citep{MorenoSarbach2003,NomuraYoshidaSoda2020,Chaverra2016}. As a result, the odd-parity metric perturbations, which in the Regge--Wheeler gauge read $h_{t\phi}=h_0(t,r)\sin\theta\,\partial_\theta P_\ell$ and $h_{r\phi}=h_1(t,r)\sin\theta\,\partial_\theta P_\ell$, are coupled to the electromagnetic perturbation $\delta A_\mu=(u_1P_\ell,\,u_2P_\ell,\,u_3\partial_\theta P_\ell,\,0)$. The scalar perturbation, being of even parity, does not enter. In this sector $\delta F\equiv0$, so neither $\Lag_{FF}$ nor $\Lag_{F\varphi}$ contributes, and the only background function that appears is $\Lf(r)=-(\rho+p_\perp)/(2F)$, which is fixed by the geometry through Eq.~\eqref{eq:identities}.

We define $\Ll=\ell(\ell+1)$, $s=\sqrt{-\Lf}$ and the tortoise coordinate $\dd r_*=e^{\lambda-\nu}\dd r$, and introduce the gauge-invariant quantities
\begin{equation}
\Theta=\partial_th_1-r^2\partial_r\!\left(\frac{h_0}{r^2}\right),\qquad
\psi=r^2e^{-(\nu+\lambda)}\Lf\,(\partial_tu_2-\partial_ru_1).
\end{equation}
Appendix~\ref{app:derivation} shows that the variables
\begin{equation}
\Psi_1=\frac1r\left[r^2e^{-(\nu+\lambda)}\Theta-\frac{8\kappa^2q}{\Ll}\,\psi\right],\qquad
\Psi_2=\frac{\sqrt{8\kappa^2(\Ll-2)}}{\Ll}\,\frac{\psi}{s},
\label{eq:masterVars}
\end{equation}
obey the symmetric system
\begin{equation}
\partial_t^2\boldsymbol\Psi-\partial_{r_*}^2\boldsymbol\Psi+\mathbf V\boldsymbol\Psi=0,
\label{eq:master}
\end{equation}
with
\begin{align}
V_{11}&=e^{2\nu}\left[\frac{\Ll}{r^2}-\frac{6m}{r^3}+\frac{\kappa^2}{2}(\rho-p_r)\right],\label{eq:V11}\\
V_{22}&=e^{2\nu}\left[\frac{\Ll}{r^2}+\frac{8\kappa^2q^2s^2}{r^4}\right]+s\,\partial_{r_*}^2\!\left(\frac1s\right),\label{eq:V22}\\
V_{12}&=V_{21}=e^{2\nu}\,\frac{q\,s}{r^3}\sqrt{8\kappa^2(\Ll-2)} .\label{eq:V12}
\end{align}
The system is valid for any $\Lag(F,\varphi)$ and any scalar background. Three remarks are in order. (i) Because $s=\sqrt{-\Lf}$ enters Eq.~\eqref{eq:masterVars}, the master variables are real only when $\Lf<0$; the axial problem is thus well defined precisely under the no-ghost condition of Section~\ref{sec:model}. (ii) The off-diagonal term carries the factor $\sqrt{(\ell-1)(\ell+2)}$, so for $\ell=1$, which does not radiate, the two channels separate. (iii) In the Maxwell limit ($s^2=1/4$, $\varphi=0$, $Q_G^2\equiv\kappa^2q^2/2$), Eqs.~\eqref{eq:V11}--\eqref{eq:V12} reduce to
\begin{equation}
\begin{gathered}
V_{11}=e^{2\nu}\!\left[\frac{\Ll}{r^2}-\frac{6M}{r^3}+\frac{4Q_G^2}{r^4}\right],\quad
V_{22}=e^{2\nu}\!\left[\frac{\Ll}{r^2}+\frac{4Q_G^2}{r^4}\right],\\
V_{12}=\frac{2e^{2\nu}Q_G\sqrt{\Ll-2}}{r^3},
\end{gathered}
\end{equation}
that is, to the Moncrief--Chandrasekhar system for RN~\citep{Moncrief1974,Chandrasekhar1983}, whose diagonalisation yields Chandrasekhar's decoupled potentials
\begin{equation}
V_j=e^{2\nu}\left[\frac{\Ll}{r^2}-\frac{q_{j}}{r^3}+\frac{4Q_G^2}{r^4}\right],\qquad q_{1,2}=3M\pm\sqrt{9M^2+4Q_G^2(\Ll-2)}.
\end{equation}
For the background considered here we also tested Eqs.~\eqref{eq:master}--\eqref{eq:V12} independently. After solving them numerically, we recovered $h_0,h_1,u_1,u_2$ from Eq.~\eqref{eq:masterVars} and substituted them into the full linearised Einstein and NED equations. Each of the six nontrivial components is satisfied with relative accuracy $10^{-11}$; for comparison, altering $V_{12}$ by 10\% raises the residuals to about $10^{-2}$. For this test we used an independent Python/SymPy implementation rather than the ancillary Mathematica program.

\subsection{Positive potential and absence of unstable modes}

\begin{figure}[htbp]
\centering
\includegraphics[width=0.72\textwidth]{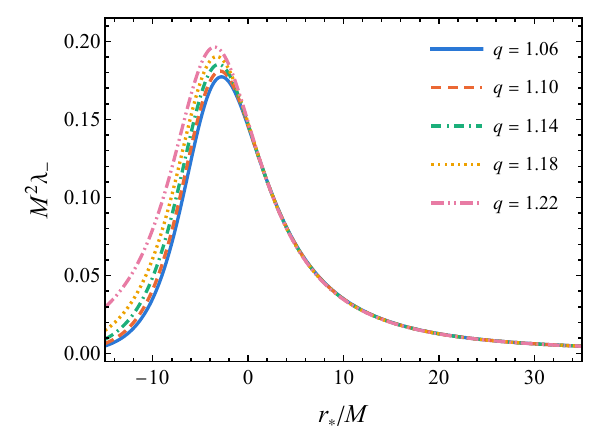}
\caption{Smallest eigenvalue $\lambda_-$ of $\mathbf V$ versus $r_*$ ($\ell=2$, $\beta=0.1$, fiducial charges). It stays above zero everywhere outside the black hole.}
\label{fig:lambda}
\end{figure}

Because $\mathbf V$ is real and symmetric, $-\partial_{r_*}^2+\mathbf V$ is a self-adjoint operator on $L^2(\mathbb R,\dd r_*)^2$. Let
\begin{equation}
\lambda_-(r)=\tfrac12(V_{11}+V_{22})-\sqrt{\tfrac14(V_{11}-V_{22})^2+V_{12}^2}
\end{equation}
denote the lower eigenvalue of the potential matrix. If $\lambda_->0$ for $r>\rp$, there are no normalisable solutions with $\omega^2<0$, so axial modes cannot grow. The function $\lambda_-$ for the fiducial charges is plotted in Fig.~\ref{fig:lambda}. Scanning the entire ghost-free window for $0\le\beta\le0.3$ and $\ell=2,\dots,6$, we find
\begin{equation}
\min_{r>\rp}\ \frac{r^2\lambda_-(r)}{e^{2\nu}}\ \ge\ 2.0 ,
\end{equation}
so that $\lambda_->0$ at all exterior radii. \emph{Hence all ghost-free black holes of the family are stable against axial perturbations.} As shown in Section~\ref{sec:radial}, this conclusion does not extend to the radial sector. The term $s\,\partial^2_{r_*}(1/s)$ in $V_{22}$ has no definite sign, but in the exterior the centrifugal and electromagnetic terms always dominate it.

\subsection{Quasinormal spectrum}

We determine the QNMs with two independent methods, described in Appendix~\ref{app:numerics}. The first one is Chebyshev collocation for the boundary-value problem in the frequency domain, with ingoing conditions at $\rp$ and outgoing conditions at infinity. The second one integrates Eq.~\eqref{eq:master} in the time domain and extracts the frequencies with the matrix-pencil method~\citep{HuaSarkar1990}. As a test, the spectral code returns the Schwarzschild frequencies~\citep{Leaver1985} with all six quoted digits and, applied to the coupled two-channel equations, yields the known RN values. For the present family, runs with two different resolutions agree to better than $10^{-7}$.

\begin{figure}[htbp]
\centering
\includegraphics[width=0.72\textwidth]{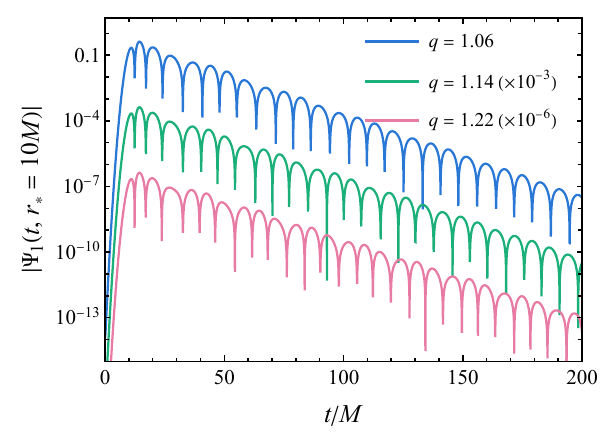}
\caption{Gravitational component $|\Psi_1|$ recorded at $r_*=10M$ for $\beta=0.1$ and $\ell=2$, starting from a Gaussian packet in $\Psi_1$. For clarity, the signals for $q=1.14$ and $q=1.22$ are shifted down by factors of $10^{3}$ and $10^{6}$. Each signal combines the two damped branches of Table~\ref{tab:qnm}, and no growing mode is present.}
\label{fig:td}
\end{figure}

\begin{figure*}[htbp]
\centering
\includegraphics[width=\textwidth]{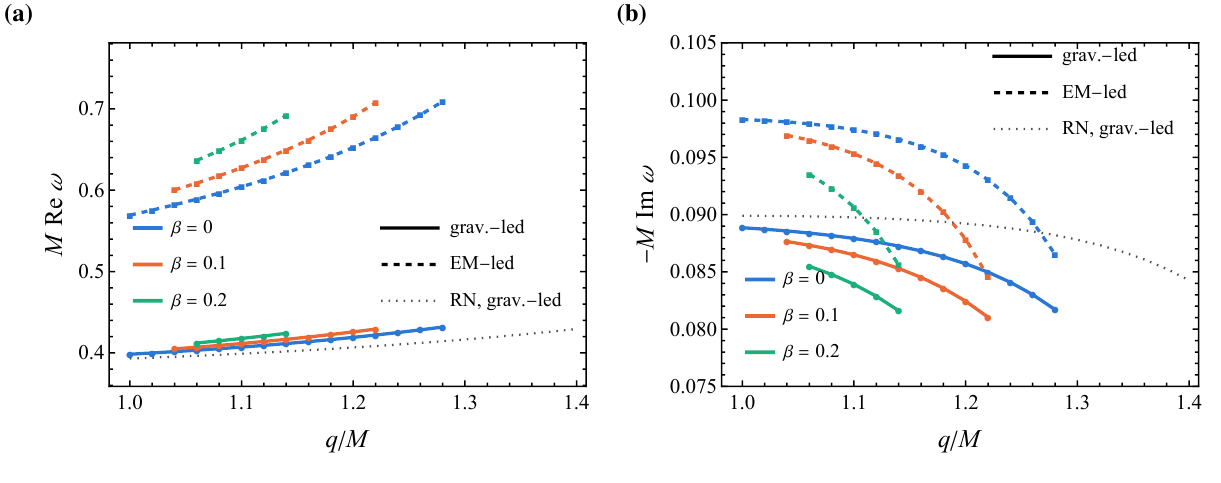}
\caption{The $\ell=2$ fundamental axial mode as a function of $q$ inside the ghost-free window, for $\beta=0,0.1,0.2$: (a) ${\rm Re}\,\omega$; (b) $-{\rm Im}\,\omega$. Solid lines: gravitational-led branch; dashed lines: electromagnetic-led branch; dotted lines: gravitational-led RN mode with the same $(M,q)$.}
\label{fig:qnm}
\end{figure*}

Two families of modes appear. One is gravity-dominated and goes over continuously into the gravitational mode of RN; the other is dominated by the electromagnetic field. Table~\ref{tab:qnm} and Fig.~\ref{fig:qnm} present the fundamental modes, and Fig.~\ref{fig:td} shows typical time-domain signals. As the positivity argument requires, all modes decay. Compared with RN of the same $(M,q)$, the gravitational-led mode oscillates 2.4--5.1\% faster and is damped 2.5--9.0\% more weakly across the $\beta=0.1$ window ($1.04\le q\le1.22$). The electromagnetic-led mode is shifted much more strongly, by 10--18\% in ${\rm Re}\,\omega$ for $\ell=2$ (9--15\% for $\ell=3$), since it propagates through the $\Lf$ profile that carries the information on the core. Both deviations increase with $\beta$ (Fig.~\ref{fig:qnm}). For the three charges evolved in time, the matrix-pencil frequencies match the spectral ones to 0.13\% (0.6\%) in the real (imaginary) part of the gravitational-led mode and to 0.3\% (1.6\%) for the subdominant electromagnetic-led mode.

\begin{table}[t]
\centering
\caption{Fundamental axial QNM frequencies $M\omega$ at $\beta=0.1$ from the spectral method (every digit shown is significant), with the RN values for equal $(M,q)$ for comparison. The labels g and em denote the gravitational-led and the electromagnetic-led branches. The bottom three rows give $\ell=2$ values extracted from time-domain (TD) signals.}
\label{tab:qnm}
\resizebox{\textwidth}{!}{%
\begin{tabular}{lcccc}
\toprule
 & \multicolumn{2}{c}{$\ell=2$} & \multicolumn{2}{c}{$\ell=3$}\\
$q$ & g & em & g & em\\
\midrule
1.06 & $0.40699-0.08733i$ & $0.60880-0.09651i$ & $0.65789-0.09020i$ & $0.84748-0.09610i$\\
 & RN: $0.39645-0.08984i$ & $0.55174-0.09902i$ & $0.63897-0.09336i$ & $0.77977-0.09915i$\\
1.10 & $0.41153-0.08649i$ & $0.62755-0.09532i$ & $0.66606-0.08924i$ & $0.87026-0.09489i$\\
 & RN: $0.39902-0.08975i$ & $0.56164-0.09908i$ & $0.64359-0.09325i$ & $0.79222-0.09916i$\\
1.14 & $0.41670-0.08530i$ & $0.64940-0.09341i$ & $0.67546-0.08790i$ & $0.89666-0.09303i$\\
 & RN: $0.40185-0.08960i$ & $0.57259-0.09905i$ & $0.64870-0.09307i$ & $0.80591-0.09909i$\\
1.18 & $0.42259-0.08358i$ & $0.67542-0.09024i$ & $0.68637-0.08595i$ & $0.92792-0.09004i$\\
 & RN: $0.40497-0.08937i$ & $0.58478-0.09889i$ & $0.65434-0.09281i$ & $0.82109-0.09888i$\\
1.22 & $0.42926-0.08104i$ & $0.70742-0.08458i$ & $0.69920-0.08298i$ & $0.96612-0.08485i$\\
 & RN: $0.40843-0.08903i$ & $0.59845-0.09853i$ & $0.66061-0.09242i$ & $0.83802-0.09849i$\\
\midrule
1.06 (TD) & $0.40693-0.08743i$ & $0.60902-0.09598i$ & & \\
1.14 (TD) & $0.41653-0.08547i$ & $0.64921-0.09245i$ & & \\
1.22 (TD) & $0.42872-0.08151i$ & $0.70527-0.08322i$ & & \\
\bottomrule
\end{tabular}}
\end{table}

\section{Radial perturbations}\label{sec:radial}

In spherical symmetry only the scalar field carries dynamics, so monopole perturbations reduce to a single wave equation. They are usually the most dangerous ones, because their effective potential has no centrifugal term. For a general scalar--electromagnetic coupling, the corresponding master equation was derived in Ref.~\citep{Bronnikov2026a}, and our action~\eqref{eq:SE} with $\Lag=f(\varphi)L(F)$ belongs to that class. In the conventions of Ref.~\citep{Bronnikov2026a} and with $\kappa^2=1$, the correspondence reads $H=1$, $\psi=\varphi/\sqrt2$ and $L(\psi,F)=2\left[V(\varphi)-\Lag(F,\varphi)\right]$. We checked it by inserting our background into the static field equations of Ref.~\citep{Bronnikov2026a}, which are then satisfied with relative residuals below $10^{-9}$.

For a purely magnetic field the invariant $F$ is not perturbed. In the gauge $\delta r=0$, the perturbation $\delta\varphi=e^{i\Omega t}\,Y(r_*)/r$ obeys
\begin{equation}
\frac{\dd^2Y}{\dd r_*^2}+\left[\Omega^2-V_{\rm eff}(r)\right]Y=0,
\label{eq:radial}
\end{equation}
with
\begin{equation}
V_{\rm eff}=e^{2\nu}\left[\mu^2_{\rm eff}+2r\varphi'\left(V_{,\varphi}-\Lag_{,\varphi}\right)+r^2\varphi'^2\left(V-\Lag-\frac1{r^2}\right)\right]+e^{2(\nu-\lambda)}\,\frac{\nu'-\lambda'}{r},\qquad
\mu^2_{\rm eff}\equiv V_{,\varphi\varphi}-\Lag_{,\varphi\varphi},
\label{eq:Veffrad}
\end{equation}
where the derivatives of $\Lag$ are taken at fixed $F$. For $\varphi\equiv0$ on a Schwarzschild background, Eq.~\eqref{eq:Veffrad} reduces to the familiar $\ell=0$ potential $2Me^{2\nu}/r^3$, and for a free field of mass $\mu$ the first term gives the usual contribution $\mu^2e^{2\nu}$. A mode with $\Omega=-i\Gamma$, $\Gamma>0$, that vanishes both at the horizon and at spatial infinity grows as $e^{\Gamma t}$ and signals an instability.

\begin{figure*}[htbp]
\centering
\includegraphics[width=\textwidth]{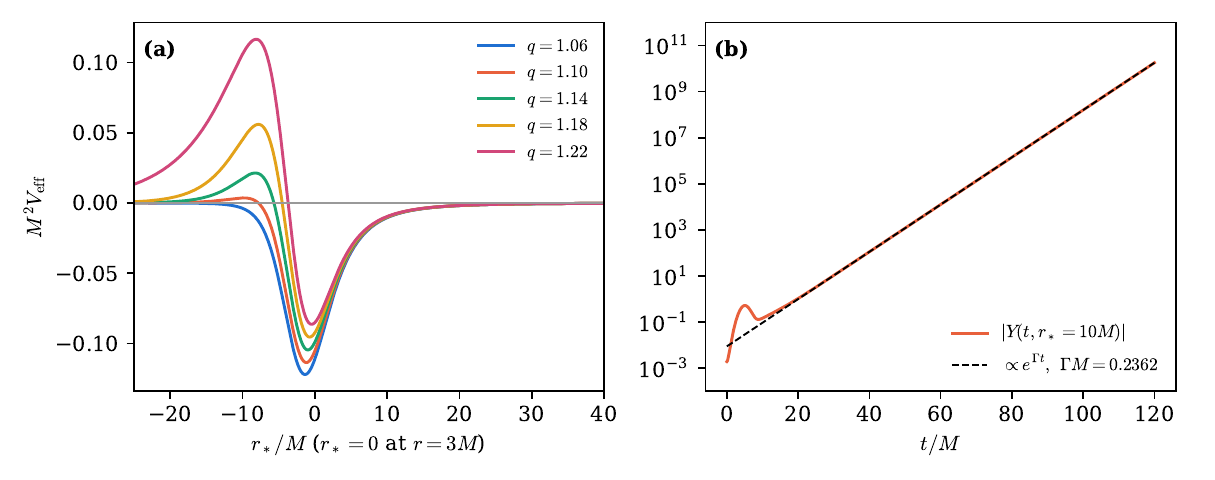}
\caption{(a) Effective potential~\eqref{eq:Veffrad} of radial perturbations for $\beta=0.1$ and the charges of Table~\ref{tab:background}, as a function of the tortoise coordinate ($r_*=0$ at $r=3M$). (b) Time evolution of $|Y|$ at $r_*=10M$ for $q=1.10$, $\beta=0.1$, starting from a Gaussian packet; the dashed line is $e^{\Gamma t}$ with the growth rate $\Gamma M=0.2362$ obtained by shooting.}
\label{fig:radial}
\end{figure*}

The potential~\eqref{eq:Veffrad} for the fiducial configurations is shown in Fig.~\ref{fig:radial}(a). It vanishes at the horizon, has a small positive barrier for charges close to $q_{\rm ext}$, and develops a deep negative well, $M^2V_{\rm eff}\approx-0.09$ to $-0.12$, at $r\approx2.5$--$2.8\,M$, followed by a tail $V_{\rm eff}\propto-r^{-3}$. The well comes from the effective mass term. Since $V\simeq-c\,\varphi^5$ with $c>0$ near $\varphi=0$ (Section~\ref{sec:model}), we have $V_{,\varphi\varphi}\simeq-20c\,\varphi^3<0$, whereas the electromagnetic contribution $\Lag_{,\varphi\varphi}$ is suppressed by $F\propto r^{-4}$. The scalar is therefore tachyonic, $\mu^2_{\rm eff}<0$, everywhere outside the horizon.

To decide whether this well supports growing modes, we used three independent procedures, described in Appendix~\ref{app:numerics}. First, the zero-energy solution of Eq.~\eqref{eq:radial} that is regular at the horizon has one or two nodes outside it, which by Sturm's oscillation theorem means one or two bound states with $\Omega^2<0$. Second, a shooting method for solutions decaying at both ends gives the growth rates directly. Third, the time evolution of Eq.~\eqref{eq:radial} from a Gaussian packet [Fig.~\ref{fig:radial}(b)] shows exponential growth with $\Gamma M=0.23619$ for $q=1.10$, $\beta=0.1$, in agreement with the shooting value $0.23619$.

\begin{table}[t]
\centering
\caption{Unstable radial ($\ell=0$) modes of the ghost-free black holes for $\sigma=\sigma_1$ ($M=1$, $\kappa^2=1$). The columns give the outer horizon $\rp$, the minimum of $V_{\rm eff}$ outside the horizon and its location, the growth rate $\Gamma$ of the fundamental unstable mode ($\delta\varphi\propto e^{\Gamma t}$), the $e$-folding time $1/\Gamma$, and the growth rate $\Gamma_1$ of the second unstable mode where it exists. The bottom part shows the dependence on $\sigma$ for $q=1.10$, $\beta=0.1$.}
\label{tab:radial}
\small\setlength{\tabcolsep}{5pt}
\begin{tabular}{cccccccc}
\toprule
$\beta$ & $q$ & $\rp$ & $M^2\min V_{\rm eff}$ & $r_{\min}$ & $\Gamma M$ & $1/\Gamma\;[M]$ & $\Gamma_1 M$\\
\midrule
0   & 0.990  & 1.597 & $-0.139$ & 2.45 & 0.279 & 3.6 & 0.076\\
0   & 1.092  & 1.474 & $-0.120$ & 2.51 & 0.250 & 4.0 & 0.045\\
0   & 1.194  & 1.302 & $-0.097$ & 2.69 & 0.202 & 4.9 & --\\
0   & 1.296  & 0.894 & $-0.073$ & 3.00 & 0.144 & 7.0 & --\\
\midrule
0.1 & 1.037  & 1.472 & $-0.127$ & 2.48 & 0.260 & 3.8 & 0.056\\
0.1 & 1.100  & 1.367 & $-0.114$ & 2.56 & 0.236 & 4.2 & 0.025\\
0.1 & 1.163  & 1.225 & $-0.099$ & 2.67 & 0.205 & 4.9 & --\\
0.1 & 1.226  & 0.912 & $-0.085$ & 2.83 & 0.169 & 5.9 & --\\
\midrule
0.2 & 1.060  & 1.316 & $-0.114$ & 2.55 & 0.235 & 4.2 & 0.021\\
0.2 & 1.091  & 1.238 & $-0.107$ & 2.60 & 0.220 & 4.5 & 0.004\\
0.2 & 1.122  & 1.134 & $-0.100$ & 2.66 & 0.204 & 4.9 & --\\
0.2 & 1.153  & 0.915 & $-0.093$ & 2.73 & 0.186 & 5.4 & --\\
\midrule
0.3 & 1.0605 & 1.071 & $-0.102$ & 2.64 & 0.207 & 4.8 & --\\
0.3 & 1.0655 & 1.038 & $-0.101$ & 2.65 & 0.204 & 4.9 & --\\
\midrule
\multicolumn{2}{c}{$\sigma/\sigma_1$} & 0.1 & 0.5 & 1 & 2 & 5 & \\
\multicolumn{2}{c}{$\Gamma M$} & 0.2357 & 0.2359 & 0.2362 & 0.2367 & 0.2382 & \\
\bottomrule
\end{tabular}
\end{table}

The results are collected in Table~\ref{tab:radial}. \emph{None of the ghost-free black holes considered here survives radial perturbations: each of them possesses a growing mode.} The fundamental growth rate lies in the range $\Gamma M\approx0.14$--$0.28$, corresponding to $e$-folding times of $3.6$--$7\,M$, and decreases towards extremality. For the smaller charges in each window a second unstable mode is present. The dependence on the NED scale is weak: varying $\sigma$ by a factor of $50$ changes $\Gamma$ by less than $1\%$, because $\sigma$ enters $V_{\rm eff}$ only through the subleading electromagnetic terms. The growth rates exceed the damping rates of the axial QNMs of Section~\ref{sec:axial}, $|{\rm Im}\,\omega|M\approx0.08$--$0.10$, so the radial instability develops faster than the axial ringdown decays. The axial spectrum therefore characterises these solutions as linear backgrounds, but not as end states of evolution.

Two remarks put this result in context. First, the conclusion is formulated in the Einstein frame. It holds in the Jordan frame of any STT whose conformal factor is smooth and positive at the horizon and at spatial infinity, since the boundary conditions are then unchanged~\citep{Bronnikov2026a}. Second, the instability is a direct consequence of the reconstruction. The profiles~\eqref{eq:ansatz} fix $V(\varphi)$, and the resulting potential has negative curvature at the asymptotic vacuum. Inverse constructions should therefore impose, in addition to regularity and ghost freedom, the requirement $\mu^2_{\rm eff}\ge0$ outside the horizon, or at least near $\varphi=0$. Whether this can be combined with an AdS core and a nonempty ghost-free window is an open question. The nonlinear end state of the instability also remains to be determined.

\section{Central instability and thermodynamics}\label{sec:polar}

\begin{figure}[htbp]
\centering
\includegraphics[width=0.72\textwidth]{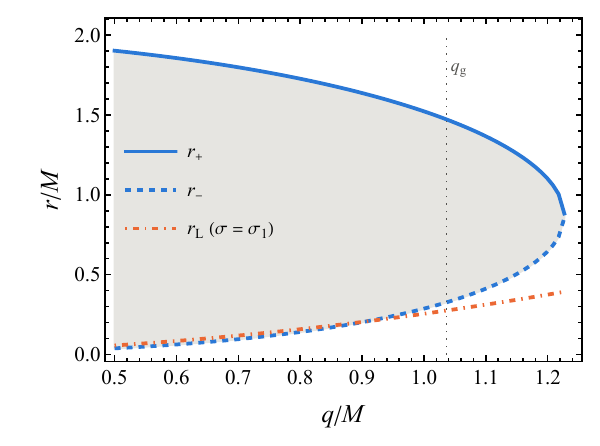}
\caption{Inner and outer horizons $r_\pm$ (shaded band between them) and the radius $r_L$ bounding the Laplacian-unstable zone, plotted against $q$ for $\beta=0.1$ and $\sigma=\sigma_1$. The dotted vertical line is at $q_{\rm g}$; to its right $r_L$ always stays below $\rmn$.}
\label{fig:radii}
\end{figure}

\paragraph{Angular instability near the centre}
The remaining sector combines polar metric perturbations, fluctuations of $\varphi$ and the axial electromagnetic mode $\delta A_\phi\propto\sin\theta\,\partial_\theta P_\ell$. Since this mode has $\delta F\neq0$, the second derivative $\Lag_{FF}$ now contributes. In the high-$\ell$ limit the highest-derivative terms of the electromagnetic equation determine the squared speed of propagation in the angular direction~\citep{DeFeliceTsujikawa2025PRL,DeFeliceTsujikawa2025},
\begin{equation}
c_\theta^2=\frac{\Lf+2F\Lag_{FF}}{\Lf}=\frac{L_F+2FL_{FF}}{L_F}=\frac{4-3\sigma F}{4+\sigma F},
\label{eq:ctheta}
\end{equation}
which is independent of $f(\varphi)$. It turns negative where $\sigma F>4/3$, i.e.\ inside the radius
\begin{equation}
r_L=\left(\tfrac32\sigma q^2\right)^{1/4}.
\end{equation}
More generally, any $L(F)$ with the asymptotics $L_c+c_1F^{-n}$ gives $c_\theta^2\to-(2n+1)$ at the centre, which agrees with the general results of Refs.~\citep{DeFeliceTsujikawa2025PRL,DeFeliceTsujikawa2025}. It is also in line with the causality theorems of Russo and Townsend~\citep{RussoTownsend2024,RussoTownsend2026}: in pure Einstein--NED theory a regular centre cannot be supported by a causal model, and in the present setting the corresponding failure of causality in the NED sector appears as $c_\theta^2<0$ for $r<r_L$. The scalar coupling does not remove this feature, because $c_\theta^2$ is independent of $f(\varphi)$. Regular magnetic cores of this type are therefore inevitably unstable against angular perturbations with large $\ell$; what distinguishes one configuration from another is only whether the unstable region can be reached causally. Since the geometry does not depend on $\sigma$ while $r_L$ does, the unstable region can be placed inside the inner horizon, $r_L<\rmn$, provided that $\sigma<\sigma_{\max}=2\rmn^4/(3q^2)$. With the choice $\sigma=\sigma_1$ this condition holds throughout the $\beta=0.1$ window (Table~\ref{tab:background} and Fig.~\ref{fig:radii}). Without horizons nothing shields the unstable zone, and such configurations must be discarded. We emphasise, however, that confining the unstable zone inside $\rmn$ does not remove the instability. The region $r<\rmn$ is static, and for $c_\theta^2<0$ the growth rate of perturbations increases without bound with $\ell$, so the initial-value problem for the NED perturbations is ill-posed there. Placing $r_L$ inside the inner horizon only makes this region causally disconnected from the exterior; it does not render the interior a viable classical configuration. The emerging picture, in which the exterior of the black hole is stable whereas an intrinsically unstable zone sits deep inside, agrees with the general conclusion drawn in Ref.~\citep{Bronnikov2026a} from combining exterior stability results for magnetic NED black holes~\citep{NomuraYoshidaSoda2020} with the central Laplacian instability~\citep{DeFeliceTsujikawa2025}. In any case, the inner horizon is expected to be subject to mass inflation~\citep{PoissonIsrael1990}. The coupled polar problem, which involves three dynamical fields, deserves a separate study.

\begin{figure}[htbp]
\centering
\includegraphics[width=0.72\textwidth]{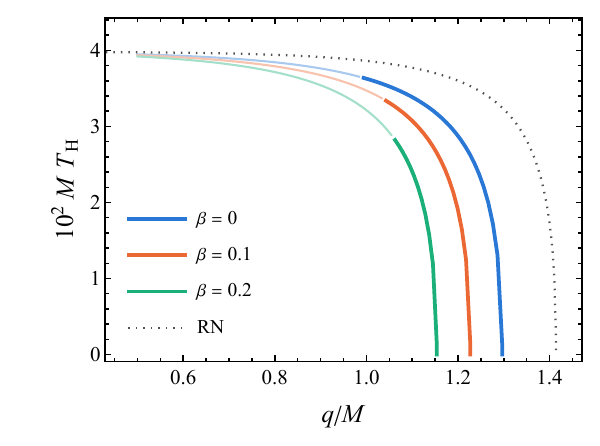}
\caption{Hawking temperature versus $q$ for $\beta=0,0.1,0.2$, compared with RN. Lighter segments mark $q<q_{\rm g}(\beta)$, where the coupling $f$ changes sign at some radius.}
\label{fig:temp}
\end{figure}

\paragraph{Hawking temperature}
From the surface gravity of the Killing vector $\partial_t$, normalised at spatial infinity, we obtain
\begin{equation}
T_H=\frac{1}{4\pi}\left(e^{-2\lambda}\right)'e^{\nu+\lambda}\Big|_{\rp}.
\end{equation}
As Fig.~\ref{fig:temp} shows, $T_H$ decreases monotonically with increasing $q$ and goes to zero at $q_{\rm ext}(\beta)$, where the two horizons coalesce. For small $q$ it approaches the Schwarzschild value $1/(8\pi M)$. At equal $(M,q)$ it is always lower than the RN temperature, partly because of the redshift factor $e^{\nu+\lambda}<1$. This comparison should not be read as an evaporation sequence. The reconstructed $f(\varphi)$ and $V(\varphi)$ depend on $(q,\beta)$, and the mass is fixed by Eq.~\eqref{eq:Mvalue}, so different points of the curves in Fig.~\ref{fig:temp} belong to different theories. Describing the evolution of a black hole under Hawking emission within a fixed theory would require time-dependent solutions, which lie beyond the present work.

\section{Summary and outlook}\label{sec:concl}

In this paper, regular black holes have been obtained in scalar--tensor gravity for the case in which a Kruglov-type NED interacts with the scalar through a coupling function $f(\varphi)$. The solutions are static, asymptotically flat and depend on two parameters. Since $\varphi$ is canonical, the AdS interior with bounded curvature is generated solely by the negative value of $V$ near $r=0$. Two general lessons for inverse constructions follow. One is that $m(0)=0$ removes the freedom in the ADM mass, which becomes a function of the charges and of the core radius. The other is that $\rho_2=8\varphi_2^2$ must hold; without this balance between the central growth of $\rho$ and the scalar kinetic energy, $f$ is singular at the origin. Requiring in addition that the photon not be a ghost leaves only the interval $q_{\rm g}(\beta)<q<q_{\rm ext}(\beta)$ for black holes, and this interval shrinks to nothing at $\beta_*\simeq0.32$.

On the perturbative side, we have shown that odd-parity metric perturbations and the electromagnetic modes they mix with obey a symmetric system of two wave equations, derived for an arbitrary $\Lag(F,\varphi)$ on a magnetically charged scalar background. This system goes over into the Moncrief--Chandrasekhar equations when the RN limit is taken and passes a direct check against the complete linearised equations. Because its potential matrix has positive eigenvalues everywhere outside $\rp$ for all admissible charges, no unstable axial modes exist. The radial sector, however, is unstable. Applying the general effective potential of monopole perturbations~\citep{Bronnikov2026a}, we found at least one growing mode for each admissible black hole, with $\Gamma M\approx0.14$--$0.28$, confirmed by shooting, by node counting and by time evolution. The cause is the reconstructed potential, $V\simeq-c\,\varphi^5$ near $\varphi=0$, which makes the scalar tachyonic outside the horizon. The present family is therefore not a set of viable end states, and inverse constructions should include the condition $\mu^2_{\rm eff}\ge0$ in the exterior as a third requirement, alongside regularity and ghost freedom. Among the lowest QNMs, which we computed spectrally and validated with time-domain runs, the gravity-dominated branch differs from RN by only a few per cent, whereas the electromagnetic one is displaced by 10--18\% at $\ell=2$; spectroscopy of the latter would therefore be the most direct way to probe the core in a radially stable variant of the construction. In the remaining sector we encounter the Laplacian instability in the angular direction that is generic for regular magnetic NED cores. It is confined to $r<r_L=(3\sigma q^2/2)^{1/4}$, which can lie under the inner horizon of a black hole but is visible from outside in horizonless solutions, so the latter are not viable. In black holes the instability becomes causally disconnected from the exterior but is not removed.

Several extensions suggest themselves. The most pressing one is to repeat the construction with profiles $\rho(r)$ and $\varphi(r)$ chosen so that the effective scalar mass is nonnegative outside the horizon, and to find out whether AdS cores and a ghost-free window survive this requirement. The nonlinear evolution of the radial instability found here, and the nonradial polar sector ($\ell\ge1$), should also be analysed. Spinning and NUT-charged versions of the solutions could be sought, perhaps with the scalar also coupled to the Gauss--Bonnet term, as was done for wormholes~\citep{Ibadov2020,Ibadov2021,Ibadov2022}. It would also be interesting to understand how these black holes are related to scalarised wormholes and to non-Abelian configurations~\citep{Hauser2014}. Finally, since photons propagate in an effective geometry that depends on $f(\varphi)$, the shadow and lensing signatures deserve study~\citep{Ramadhan2026,DeParaAzregAinou2025}. Since the scalar charge modifies $g_{rr}$ and the lapse already at $O(r^{-2})$, the circular geodesics, and hence the properties of thin accretion disks, should also differ from the RN case, as found for other regular black holes~\citep{Akbarieh2025,Khoshrangbaf2025}.

\section*{Acknowledgements}
The authors thank Kirill Bronnikov for helpful remarks and fruitful discussions. Funding for this study was received from the Ministry of Higher Education, Science and Innovations of the Republic of Uzbekistan through project No.~FZ-20200929385.

\section*{Data and code availability}
Every figure except Fig.~\ref{fig:radial}, and every entry of Tables~\ref{tab:background} and~\ref{tab:qnm}, can be regenerated with one Wolfram Mathematica script, \texttt{RBH\_figures.wl} (version 13.0), supplied as an ancillary file and described in Appendix~\ref{app:numerics}. The cross-check of the axial equations against the full linearised system in Section~\ref{sec:axial} relied on independent Python/SymPy code, and the radial-perturbation analysis of Section~\ref{sec:radial} (Table~\ref{tab:radial} and Fig.~\ref{fig:radial}) on the Python/NumPy script \texttt{radial\_perturbations.py}, which is also supplied as an ancillary file. One complete execution of the script needs roughly four minutes on an ordinary PC; it writes out the plots, the \LaTeX{} sources of the tables, and the underlying numbers as \texttt{.csv} and \texttt{.m} files.

\appendix

\section{Derivation of the two-channel axial equations}\label{app:derivation}

We linearise Eqs.~\eqref{eq:Einstein}--\eqref{eq:scalarMaxwell} with the ansatz of Section~\ref{sec:axial}, using computer algebra for general $\nu,\lambda,\varphi,V,\Lag$, and separate the angular dependence. This leaves six nontrivial equations. The $(\theta\phi)$ component of the Einstein equations gives
\begin{equation}
\partial_th_0=e^{\nu-\lambda}\partial_r\!\left(e^{\nu-\lambda}h_1\right).
\end{equation}
In the $(r\phi)$ component the background terms combine into $-4\kappa^2F\Lf$ because of Eq.~\eqref{eq:fluid}. With the notation $\mathcal{B}\equiv u_2-\partial_ru_3$ this component becomes
\begin{equation}
r^2e^{-2\nu}\partial_t\Theta=-\left(\Ll-2-\frac{8\kappa^2q^2\Lf}{r^2}\right)h_1-8\kappa^2q\Lf\,\mathcal{B} .
\end{equation}
For the NED field equations, the radial and temporal components give
\begin{equation}
\partial_t\psi=\frac{\Ll\Lf e^{\nu-\lambda}}{r^2}\left(qh_1-r^2\mathcal{B}\right),\qquad
\partial_r\psi=-\frac{\Ll\Lf e^{\lambda-\nu}}{r^2}\left(r^2\mathcal{A}-qh_0\right),
\end{equation}
where $\mathcal{A}\equiv u_1-\partial_tu_3$. The $\theta$ component and the $(t\phi)$ equation follow from these relations. Eliminating $\mathcal{B}$, we get $r^2e^{-2\nu}\partial_t\Theta=-(\Ll-2)h_1+(8\kappa^2q/\Ll)e^{\lambda-\nu}\partial_t\psi$; combining this with the $(\theta\phi)$ relation and integrating once in time, we arrive at
\begin{equation}
\left[\partial_t^2-r^2\partial_{r_*}\frac1{r^2}\partial_{r_*}\right]\!\left(\Pi-\frac{8\kappa^2q}{\Ll}\psi\right)+\frac{(\Ll-2)e^{2\nu}}{r^2}\Pi=0,
\end{equation}
where $\Pi\equiv r^2e^{-(\nu+\lambda)}\Theta$. Eliminating $\mathcal{A}$ and $\mathcal{B}$ between $\partial_t\psi$ and $\partial_r\psi$ gives
\begin{equation}
\partial_t^2\psi-\Lf\partial_{r_*}\!\left(\frac{\partial_{r_*}\psi}{\Lf}\right)+\frac{\Ll e^{2\nu}}{r^2}\psi-\frac{\Ll q\Lf e^{2\nu}}{r^4}\Pi=0 .
\end{equation}
The first-derivative terms disappear after the substitution $\Pi=r\Psi_1+(8\kappa^2q/\Ll)\,s\,\gamma\Psi_2$, $\psi=s\gamma\Psi_2$, where the constant $\gamma=\Ll/\sqrt{8\kappa^2(\Ll-2)}$ is chosen so that the coupling becomes symmetric. Using $\partial_{r_*}^2r/r=e^{2(\nu-\lambda)}(\nu'-\lambda')/r$ and the background identity $e^{-2\lambda}(\nu'-\lambda')/r=2m/r^3+\kappa^2(p_r-\rho)/2$, we obtain Eqs.~\eqref{eq:master}--\eqref{eq:V12}.

\section{Details of the numerical work}\label{app:numerics}

The script \texttt{RBH\_figures.wl} (Mathematica 13.0) produces all plotted curves except those of Fig.~\ref{fig:radial}, every tabulated value and the energy-condition percentages quoted in Section~\ref{sec:model}. It proceeds in eight steps. It first defines the model symbolically and verifies the field equations. Next it determines $q_{\rm ext}(\beta)$, $q_{\rm g}(\beta)$ and $\beta_*$, fills Table~\ref{tab:background} and tests positivity of the axial potential. It then runs the spectral QNM code, performs the time evolution together with the temperature scans, and finally draws the plots and writes the \LaTeX{} tables.

\paragraph{Background solution and its verification}
Exact symbolic expressions in $(r,a,C,\beta)$ are built for the mass function, the metric functions, $\varphi$, $\rho$, $p_r$, $p_\perp$, $R$ and $R_{\mu\nu\alpha\beta}R^{\mu\nu\alpha\beta}$, and are then compiled (\texttt{Compile}, listable) for fast evaluation. The code verifies symbolically that $m'=r^2\rho/2$, $m(\infty)=\pi C(3-\beta)/(64a)$ and $(\nu+\lambda)'=r\varphi'^2/2$, and that near $r=0$ one has $\rho+p_\perp=O(r^4)$ with the coefficient $C[36a^2(1-\beta)+7C\beta(1+3\beta)]/(48a^{10})>0$, which determines $\sigma_1$. After the reconstructed $f(\varphi)$ and $V(\varphi)$ are substituted back, the $tt$, $rr$, $\theta\theta$ components of the Einstein equations and the Klein--Gordon equation hold with residuals smaller than $3\times10^{-12}$ on $0.05\le r/M\le5$ ($q=1.10$, $\beta=0.1$). To locate the horizons, \texttt{FindRoot} is applied in each interval where $e^{-2\lambda}$ changes sign. Brent's method gives $q_{\rm ext}(\beta)$ from the condition $\min_r e^{-2\lambda}=0$ and $q_{\rm g}(\beta)$ from $\min_r(\rho+p_\perp)=0$, and $\beta_*=0.322538$ is the crossing point of the two curves.

\paragraph{Frequency-domain solver}
With $N\equiv e^{\nu-\lambda}=\dd r/\dd r_*$ and the surface gravity $\kappa_+=N'(\rp)/2$, we write $\boldsymbol\Psi=e^{-i\omega t}e^{i\omega H(r)}\mathbf Z(u)$, where $u=\rp/r$ and
\begin{equation}
H'(r)=\frac1N-\frac{\rp}{\kappa_+r(r-\rp)} .
\end{equation}
Near the horizon the factor $e^{i\omega H}$ behaves like $e^{-i\omega r_*}$, and at infinity like $e^{+i\omega r_*}$; therefore $\mathbf Z$ is regular at both ends. The equation for $\mathbf Z$,
\begin{multline}
N^2\mathbf Z''+NN'\mathbf Z'-\mathbf V\mathbf Z+i\omega\left[2N^2H'\mathbf Z'+N(NH''+N'H')\mathbf Z\right]\\
+\omega^2(1-N^2H'^2)\mathbf Z=0,
\end{multline}
is multiplied by $r^2$ and collocated at $n$ Chebyshev--Gauss points in $u\in(0,1)$; the derivative matrices are constructed from the polynomials $T_j$ and $U_j$. To prevent loss of precision we rearrange the coefficients so that no cancellations occur; for example, $1-N^2H'^2=2\delta-\delta^2$ with $\delta=N\rp/[\kappa_+r(r-\rp)]$. This yields a quadratic eigenproblem $(A_0+\omega A_1+\omega^2A_2)\mathbf Z=0$, which we linearise via the companion matrix and solve as a dense generalised eigenproblem with \texttt{Eigenvalues[}$\{A,B\}$\texttt{]}~\citep{Jansen2017}. We keep a mode only when the frequencies obtained with $n=44$ and $n=60$ coincide to $10^{-7}$ (for the Schwarzschild and RN benchmarks, $n=40$ and $56$). In the Schwarzschild case we obtain $M\omega=0.373672-0.0889623i$ ($\ell=2$) and $0.599443-0.0927030i$ ($\ell=3$), matching Leaver~\citep{Leaver1985}.

\paragraph{Time evolution}
The function $r(r_*)$ is found by integrating $\dd r/\dd r_*=N(r)$ with \texttt{NDSolveValue} (both tolerance goals set to $10^{-12}$), normalised by $r(r_*=0)=4M$. Equation~\eqref{eq:master} is advanced in time on a uniform $r_*$ grid, with centred fourth-order spatial differences implemented through \texttt{ListCorrelate}, a classical fourth-order Runge--Kutta integrator and packed arrays ($\Delta r_*=0.04M$, $\Delta t=0.4\Delta r_*$) on the interval $r_*\in[-160,260]M$. For this domain, waves reflected from the boundaries reach the observer at $r_*=10M$ only after $t=200M$. As initial data we put a Gaussian of width $1.5M$ into $\Psi_1$ at $r_*=-2M$, leaving $\Psi_2=0$. Frequencies are fitted on $35M\le t\le90M$ by the matrix-pencil technique~\citep{HuaSarkar1990}, using \texttt{SingularValueDecomposition} and keeping six singular values. This procedure produced Fig.~\ref{fig:td} and the TD entries of Table~\ref{tab:qnm}.

\paragraph{Radial perturbations}
The derivatives $V_{,\varphi}$, $V_{,\varphi\varphi}$, $\Lag_{,\varphi}$ and $\Lag_{,\varphi\varphi}$ entering Eq.~\eqref{eq:Veffrad} are obtained from the reconstructed functions of $r$ by the chain rule, using fourth-order finite differences with step $2\times10^{-3}r$ (the derivative $p_r'$ inside $p_\perp$ is computed by the complex-step method). Node counting uses the zero-energy solution of Eq.~\eqref{eq:radial} integrated with a fourth-order Runge--Kutta scheme in the variable $\ln(r/\rp-1)$ from $r-\rp=10^{-7}\rp$ to $r=400\rp$. The results do not change when the outer boundary is moved to $200\rp$ or $800\rp$ or when the resolution is doubled. Shooting starts from $Y\propto e^{\Gamma r_*}$ at the horizon, and $\Gamma$ is fixed by requiring the coefficient of the growing exponential at $r=150$--$600\,\rp$ to vanish; the fundamental growth rate is stable to $10^{-7}$. The time evolution uses second-order centred differences in $r_*\in[-120,200]M$ with $\Delta r_*=0.05M$, fourth-order Runge--Kutta stepping with $\Delta t=\Delta r_*/2$, and an initial Gaussian of width $2M$ centred at $r_*=5M$. The growth rate is fitted on $60M\le t\le140M$.

\paragraph{Plotting}
Plots are produced with Mathematica's built-in graphics, except Fig.~\ref{fig:radial}, which is drawn with Matplotlib, and all are saved as vector PDFs; Figs.~\ref{fig:fV} and~\ref{fig:qnm} are each assembled from two panels of this kind. The symmetric-logarithmic axis of Fig.~\ref{fig:fV}(b) uses the map $y\mapsto\operatorname{arcsinh}(y/y_0)$ with $y_0=0.02$.

\end{document}